\documentclass[11pt]{article}
\usepackage[utf8]{inputenc}
\usepackage{amsmath}
\usepackage{amsfonts}
\usepackage{amssymb}
\usepackage{amsthm}

\def\p{\partial}

\newtheorem{prop}{Proposition}

\theoremstyle{remark}
\newtheorem*{rem}{Remark}
\newcommand{\dbar}{\bar{\partial}}
\newcommand{\wt}{\widetilde}
\newcommand{\be}{\begin{equation}}
\newcommand{\ee}{\end{equation}}
\newcommand{\bea}{\begin{eqnarray}}
\newcommand{\eea}{\end{eqnarray}}
\newcommand{\beaa}{\begin{eqnarray*}}
\newcommand{\eeaa}{\end{eqnarray*}}

\newcommand{\nn}{\nonumber}

\usepackage{authblk}
\title{Linearly degenerate hierarchies of quasiclassical SDYM type}
\author[1]{L.V. Bogdanov \thanks{leonid@itp.ac.ru}}
\author[2,3,4]{M.V. Pavlov}
\affil[1]{L.D. Landau ITP, Moscow}
\affil[2]{
Lebedev Physical Institute, Moscow
}
\affil[3]{
National Research Nuclear University MEPHI, Moscow
}
\affil[4]
{
Department of Mechanics and Mathematics,
Novosibirsk State University,
Novosibirsk, Russia
}

\date{}
\begin{document}

\maketitle
\begin{abstract} We demonstrate that
SDYM equations for the Lie algebra of one-dimensional vector
fields represent a natural reduction in the framework
of general linearly degenerate dispersionless hierarchy.
We define the reduction in terms of wave functions, introduce
generating relation,
Lax-Sato equations and the dressing scheme for the reduced hierarchy.
Multidimensional case is also discussed.
\end{abstract}
\section{Introduction}
This work is connected with the class
of equations containing quasiclassical self-dual Yang-Mills equations. Let us consider a Lax pair
\bea
&&
L=\p_{t_1}-\lambda \p_{y_1}+A_1,
\nn
\\
&&
M=\p_{t_2}-\lambda \p_{y_2}+A_2,
\label{YMpair}
\eea 
where $A_1$, $A_2$ belong to some Lie algebra,
$\lambda$ is a complex variable (spectral parameter).
Commutativity condition for operators (\ref{YMpair}) implies the
existence of potential $F$, $A_1=\p_{y_1}F$, $A_2=\p_{y_2}F$,
satisfying the equation
\bea
\p_{t_1}\p_{y_2}F-\p_{t_2} \p_{y_1}F-
[\p_{y_1}F,\p_{y_2}F]=0.
\label{YM}
\eea 
For the case of matrix Lie algebra this is a well-known
(complexified)
self-dual Yang-Mills equation (see, e.g., \cite{AT93},
\cite{DunajBook}). For the case of Lie algebra of vector fields,
$F=\sum_{i=1}^N v^i\p_{x_i}$,
equation (\ref{YM}) represents 
$(N+4)$-dimensional
`quasiclassical' self-dual Yang-Mills equations for the coefficients of vector field $F$
(see \cite{PlebPrzan96}, \cite{ManSan2006})
\bea 
\p_{t_1}\p_{y_2}v^i-\p_{t_2} \p_{y_1}v^i
- \sum_{j=1}^N (\p_{y_1}v^j)\p_{x_j}\p_{y_2}v^i
+\sum_{j=1}^N (\p_{y_2}v^j)\p_{x_j} \p_{y_1}v^i=0,
\label{quasiYM}
\eea 
where $1\leqslant i\leqslant N$,
which for two-dimensional Hamiltonian
vector fields reduces to six-dimensional heavenly equation \cite{PlebPrzan96}.

In this work we will consider mainly one-dimensional vector
fields, $F=v\p_x$, then linear equations 
corresponding to Lax pair (\ref{YMpair}) are
\bea
&&
\p_{t_1}\Psi=(\lambda \p_{y_1} - (\p_{y_1}v)\p_x)\Psi,
\nn
\\
&&
\p_{t_2}\Psi=(\lambda \p_{y_2}-(\p_{y_2}v)\p_x)\Psi,
\label{5pair}
\eea 
and compatibility conditions (\ref{quasiYM}) take the form
of one five-dimensional equation
\bea
\p_{t_1}\p_{y_2}v-\p_{t_2} \p_{y_1}v
- (\p_{y_1}v)(\p_x\p_{y_2}v)
+(\p_{y_2}v)(\p_x\p_{y_1}v)=0.
\label{5eq}
\eea 
Performing dimensional reduction $\p_{y_1}=\p_x$,
$\p_{t_1}=\p_{y_2}(=\p_y)$
(respectively, considering 
$\Psi=\Psi(\lambda, x+y_1, t_1+y_2,t_2=t)$),
we come to linear equations
\bea
&&
\p_{y}\Psi=(\lambda \p_{x} - v_x\p_x)\Psi,
\nn
\\
&&
\p_{t}\Psi=(\lambda \p_{y}-v_y\p_x)\Psi
\label{3pair}
\eea 
corresponding to (2+1)-dimensional equation \cite{PavlovEq}
\bea
v_{tx}=v_{yy}+
v_x v_{xy}-
v_y v_{xx},
\label{Pavlov0}
\eea 
which is a simple representative of a class of linearly degenerate (or weakly nonlinear)
dispersionless integrable 
equations 
(see \cite{FK}, \cite{FK1}, \cite{FKK09})), 
connected with the so-called universal
hierarchy of Shabat and Martinez Alonso \cite{UniSA}.
A more general geometric setting for the reduction of $\mbox{Diff}(S^1)$ SDYM equations
to equation (\ref{Pavlov0}) was presented in \cite{DS2005}.

Our main purpose is a description of the hierarchy connected with
equations (\ref{5pair}) and its generalizations
in the framework of
general multidimensional dispersionless hierarchy, using the technique
developed in \cite{BK}, \cite{BDM}, \cite{LVB09}. 
We will demonstrate that 
hierarchies of this
type represent a rather natural reduction of generic 
linearly degenerate dispersionless case. We will also discuss an immersion of `universal' hierarchy of Shabat and Martinez Alonso \cite{UniSA} to 
multidimensional hierarchies connected with equations of the type (\ref{5pair}), (\ref{5eq}).

A natural language to consider multiple commuting flows
or the whole hierarchy for equations connected with commuting vector fields is a language of distributions. A distribution
$\Delta$ is an $K$-dimensional subbundle of the tangent bundle
of smooth $N$-dimensional manifold, possessing  a basis of $K$
smooth vector fields $V_1,\dots V_{K}$. Taking local coordinates
$x_1,\dots, x_N$, we can write the basis as
$V_i=\sum_{k=1}^{N} v_i^k \p_k$, $\p_k=\frac{\p~}{\p x_k}$. A distribution $\Delta$ is called involutive if 
$[\Delta,\Delta]\subset \Delta$ (in the sense of commutation of vector fields). According 
to Frobenius theorem for vector fields,
for the case of involutive distributions a system of  partial differential equations 
\bea 
V_i f:= \sum_{k=1}^{N} v_i^k \p_k f=0,
\quad 
1\leqslant i\leqslant K
\label{vectoreqn}
\eea 
locally admits $M=N-K$ (we will call $M$ a codimension of the
distribution) independent solutions $f_1,\dots, f_M$.
After some renotation of variables, the basic vector fields
can always be chosen in a form
\bea
V_i=\frac{\p~}{\p t_i}-\sum_{k=1}^{M} v_i^k \p_k,
\quad 
1\leqslant i\leqslant K
\label{basisflows}
\eea
where local variables are 
now 
$t_1,\dots, t_K$, 
$x_1,\dots,x_M$. 
For the involutive (integrable) case the basic vectors
$V_i$ commute, each subset of basic vectors gives a subdistribution with the same codimension (in remaining variables), and the Jacobian of functions $f_1,\dots, f_M$
over variables $x_1,\dots, x_M$ is not equal to zero. The
distribution is tangent to the level surfaces of the
functions $f_1,\dots, f_M$ and is uniquely defined by these
functions.

The basis of the form (\ref{basisflows}) is a good starting
point to consider the system of commuting flows, where
the variables $t_1,\dots,t_K$ are `times' of the flows.
One more ingredient we need to consider dispersionless integrable equations is holomorphicity. 
We suggest that coefficients of the
basis 
$v_i^k$ depend holomorpically on the spectral variable 
$\lambda$. In the present work, we consider mostly the simplest case
of polynomial coefficients. Then commutativity (or, more generally, involutivity) of vector fields provides equations
for the coefficients of polynomials. Solutions 
of equations (\ref{vectoreqn}) $f_1,\dots, f_M$
depend in this case on the spectral parameter, and we call them
wave fuctions (using the terminology common 
in the theory of integrable equations). A general wave
function is given by the expression 
$\Psi(\lambda,\mathbf{x})=
F(\lambda,f_1(\lambda,\mathbf{x}),\dots, f_M(\lambda,\mathbf{x}))$, and the transformation between
two sets of basic wave functions is defined by $M$ functions
of $M+1$ variables. This transformation defines functional freedom 
in the Riemann-Hilbert problem (see below) and indicates that {\em  (maximal) dimensionality of integrable equations corresponding to the distribution of codimension
$M$ is $M+2$}.

Let us consider linear equations (\ref{3pair}) from this viewpoint.
These equations correspond to basic vectors of two-dimensional
distribution in three-dimensional  space  of variables
$x$, $y$, $t$. Involutivity is provided by 
equation (\ref{Pavlov0}),
and, according to Frobenius theorem, 
equations (\ref{3pair}) admit
one independent solution (wave function) $\Psi(\lambda, x,y,t)$,
and a general solution is given locally by the function
$F(\Psi,\lambda)$. The function $\Psi$ can be found
in the form of series in $\lambda$, 
\bea 
\Psi=\Psi_0+\wt\Psi,
\qquad 
\Psi_0=\lambda^2t+\lambda y+x,\quad
\wt\Psi=\sum_{n=1}^\infty \Psi_n(x,y,t)\lambda^{-n},
\label{Psi}
\eea
where the coefficients  $\Psi_n(x,y,t)$ are defined recursively through linear equations (\ref{3pair}). Linear equations (\ref{3pair}) can be also rewritten in the form corresponding
to standard basis structure (\ref{basisflows}),
\beaa 
&&
\p_{y}\Psi=(\lambda \p_{x} - v_x\p_x)\Psi,
\nn
\\
&&
\p_{t}\Psi=(\lambda^2 \p_{x}-\lambda v_x\p_x -v_y\p_x)\Psi,
\eeaa  
which can be extended to the infinite hierachy of the form
\bea
\p_{t_n}\Psi=(\lambda^n- P_{n-1}) \p_{x}\Psi,
\quad 1\leqslant n < \infty,
\label{3pairB}
\eea 
where $P_{n-1}$ are polynomials in $\lambda$ of the order
$n-1$ with the coefficients depending on $x$ and times.
Solution (\ref{Psi}) retains its structure, with $\Psi_0$
generalized to the infinite set of times, 
\bea 
\Psi=\Psi_0+\wt\Psi,
\quad 
\Psi_0=\sum_{k=0}^{\infty} \lambda^k t_k,
\quad 
\wt\Psi=\sum_{n=1}^\infty \Psi_n(\mathbf{t})\lambda^{-n},
\label{Psigen}
\eea
and $x=t_0$, $y=t_1$, $t=t_2$.
Codimension of correspondinq distribution is 1 for any restriction to the finite set of times, basic vectors for the case of involutive distribution are commuting and commutation condition for any pair of them gives a closed (2+1)-dimensional system for the coefficients of corresponding polynomials.
The hierarchy can be also written in recursive form, which 
continues the basic set of the form (\ref{3pair}),
\bea 
\p_{t_n}\Psi=(\lambda \p_{t_{n-1}}-v_{t_{n-1}}\p_x)\Psi,
\quad 1\leqslant n < \infty,\;x=t_0.
\label{3pairRec}
\eea 
This hierarchy is a simplest example of linearly-degenerate
dispersionless hierarchy, it corresponds to the so-called universal
hierarchy of Shabat and Martinez Alonso \cite{UniSA} for 
positive times. The picture of the hierarchy can be developed
starting from the wave function in the form
of series (\ref{Psigen}) and formulating the dynamics in 
terms of the series, or, equivalently,
in terms of hydrodynamic type chains \cite{PavlovEq}.

Let us proceed now to the initial SDYM equation for
the Lie algebra of one-dimensional vector fields
(five-dimensional quasiclassical SDYM equation) 
(\ref{5pair}), (\ref{5eq}).  
Codimension of the distribution is three,
and equations (\ref{5pair}) should have three 
independent solutions.
One of these solutions is of the form (\ref{Psi}),
\bea
\Psi=x+\sum_{n=1}^\infty \Psi_n(\mathbf{t})\lambda^{-n},
\label{PsiX}
\eea
and two others are trivial,
\beaa 
\Psi^1=\lambda t_1+y_1, \quad   \Psi^2=\lambda t_2+y_2.
\eeaa
In \cite{BDM}, \cite{LVB09} we developed a picture of general multidimensional dispersionless
hierarchy 
corresponding to $N$ wave functions of the form 
(\ref{Psigen}) with $N$ infinite sets of times. From this viewpoint, equation (\ref{5eq}) is connected with a special reduction of general linearly degenerate 5-dimensional 
dispersionless hierarchy, characterised by the condition $\Psi^1=\Psi^1_0$,
$\Psi^2=\Psi^2_0$. Thus two of the wave functions are analytic
(polynomial for finite set of variables). In terms of initial
Lax pair (\ref{YMpair}) 
the general case corresponds to the presence
of generic terms with derivatives  $\p_{y_1}$, $\p_{y_2}$ in
both vector fields $A_1$, $A_2$  (and then in $F$), that
corresponds to canonical structure of general basis for the distribution
(\ref{basisflows}). In the reduced case some of the coefficients of
vector fields are equal to zero, and reduction is characterised
by the presence of analytic (polynomial) wave functions.
The functional freedom in the Riemann-Hilbert problem
for the reduced case is one function of four variables (see below), that corresponds to the fact that the reduction is not dimensional,
and the reduced systems preserves the dimensionality five.

There exists an interesting direct immersion of
(2+1)-dimensional hierarchy (\ref{3pairB}),
(\ref{3pairRec}) to the five-dimensional
system (\ref{5eq}), (\ref{5pair}).  
Indeed, let us consider a pair
of linear equations (\ref{3pairRec}) 
corresponding to two distinct pairs of
times  $t_n,\; t_{n-1}$ and $t_k,\; t_{k-1}$,
\beaa 
&&
\p_{t_n}\Psi=(\lambda \p_{t_{n-1}}-v_{t_{n-1}}\p_x)\Psi,
\\
&&
\p_{t_k}\Psi=(\lambda \p_{t_{k-1}}-v_{t_{k-1}}\p_x)\Psi
\eeaa
This system exactly corresponds to linear system (\ref{5pair}).
Thus, having a dependence of solution of the 
(2+1)-dimensional hierarchy 
(\ref{3pairB}) on two arbitrary distinct pairs of higher times,
we get a special solution to five-dimensional
equation (\ref{5eq}). Thus equation
(\ref{5eq}) plays a role of 'intertwiner' 
for higher flows 
of the hierarchy associated with the distribution (\ref{3pairB}),
(\ref{3pairRec}). A similar phenomenon is known for the case of SDYM hierarchy \cite{AT93}.
\subsection{Towards general multidimensional case}
\label{Subexample}
Let us consider first a simple characteristic example which demonstrates some distinctive features of quasiclassic SDYM case, connected with the fact that corresponding distributions 
are rather special, starting from three-dimensional distribution
with the basic equations 
\bea
&&
\p_{t_1}\Psi=(\lambda \p_{y_1} - v_{y_1}\p_x)\Psi,
\nn
\\
&&
\p_{t_2}\Psi=(\lambda \p_{y_2}-v_{y_2}\p_x)\Psi,
\label{5pair3}
\\
&&
\p_{t_3}\Psi=(\lambda \p_{y_3} - v_{y_3}\p_x)\Psi.
\nn
\eea 
Codimension of this distribution is 4, and we expect integrable
systems of (maximal) dimension 6. However, taking compatibility
condition for any pair of equations (\ref{5pair3}), we
get equations of dimension 5. This phenomenon is connected with
presence of polynomial solutions $\Psi^1=\lambda t_1+y_1$, 
$\Psi^2=\lambda t_2+y_2$, $\Psi^3=\lambda t_3+y_3$ (the fourth function
of the basic set is generic, of the form (\ref{PsiX})). Lax pair for some integrable equation is obtained from the distribution 
(containing several commuting flows) by restriction to some 
subset of variables (or, more generally, by restriction to some 
submanifold) to obtain two-dimensional distribution, 
which is especially simple for the basis of the form (\ref{basisflows}).
For the case of generic wave functions, the number of wave 
functions and respectively the codimension of the distribution
is preserved, and we get $R+2$-dimensional equations.
However, if there exist special wave functions independent
on some of the variables (like in the present example), they may
drop out in the process of restriction of the set of variables,
then the codimension of the distribution and respectively the
dimensionality of the system will be less then maximal. 

One of the ways to get 5-dimensional system in this case is
a restriction of the set of variables.
Let us consider a restriction to submanifold $x=\text{const}$.
The distribution in this case is two-dimensional and consists of the vectors, not containing $\p_x$, the basis can be obtained
from (\ref{5pair3}), taking appropriate linear combinations.
In terms of the basic equations, we can express $\p_x\Psi$ from
the  first equation in (\ref{5pair3}),
\beaa 
\p_x\Psi=
-\frac{1}{v_{y_1}}
(\p_{t_1}-\lambda\p_{y_1})\Psi.
\eeaa
Substituting this expression into third and and second 
equations of the system (\ref{5pair3}), we obtain
\bea  
&&
(\p_{t_2}-\lambda \p_{y_2})\Psi=
\frac{v_{y_2}}{v_{y_1}}
(\p_{t_1}-\lambda\p_{y_1})
\Psi,
\nn
\\
&&
(\p_{t_3}- \lambda \p_{y_3})\Psi=\frac{v_{y_3}}{v_{y_1}}
(\p_{t_1}-\lambda\p_{y_1})
\Psi,
\label{5pair3A}
\eea 
the compatibility condition for this system (the commutativity
of the basic vectors of the distribution)
gives a
quasilinear equations of second order,
\begin{equation}
{v}_{y_{3}}({v}_{y_{2}t_{1}}-{v}_{y_{1}t_{2}})+
{v}_{y_{1}}({v}_{y_{3}t_{2}}-{v}_{y_{2}t_{3}})+
{v}_{y_{2}}({v}_{y_{1}t_{3}}-{v}_{y_{3}t_{1}})
=0.
\label{six}
\end{equation}

\begin{rem}
This six-dimensional quasilinear equation of the second order is connected with the two-component first order quasilinear
system \cite{ZK}
\beaa
&&
fg_{y_{1}}-g f_{y_{1}}+f_{y_{3}}-g_{y_{2}}=0,
\nn\\
&&
gf_{t_{1}}-fg_{t_{1}}+g_{t_{2}}-f_{t_{3}}=0,
\eeaa 
associated with a Lax pair
\beaa  
&&
(\p_{t_2}-\lambda \p_{y_2})\Psi=
f
(\p_{t_1}-\lambda\p_{y_1})
\Psi,
\nn
\\
&&
(\p_{t_3}- \lambda \p_{y_3})\Psi=g
(\p_{t_1}-\lambda\p_{y_1})
\Psi,
\eeaa 
linear system (\ref{5pair3A}) and equation (\ref{six}) are obtained by the substitution \cite{artur}
\beaa 
f=\frac{v_{y_2}}{v_{y_1}},\quad g=\frac{v_{y_3}}{v_{y_1}}.
\eeaa
\end{rem}

Another way is to make a restriction to a slightly more
involved submanifold, 
the advantage of this way is that it can be
easily generalized to an arbitrary number of the basic equations of the form (\ref{5pair3}). Let us consider a restriction of the distribution (\ref{5pair3})
to submanifold $y_2=t_3=\tau$. Then we get 2-dimensional
distribution in the space of 6 variables $t_1$, $y_1$,
$t_2$, $\tau$, $y_3$, $x$
with a basis
\bea
&&
\p_{t_1}\Psi=(\lambda \p_{y_1} - (\p_{y_1}v)\p_x)\Psi,
\nn
\\
&&
\p_{t_2}\Psi=(-\lambda^2\p_{y_3}+
\lambda \p_{\tau} +(\lambda(\p_{y_3}v) + u) \p_x)\Psi,
\label{5pair36}
\eea 
and a compatibility condition gives a six-dimensional
system of equations for two functions $u$, $v$,
\beaa 
{u}_{y_1} - {v}_{y_3t_1}-{v}_{y_1 \tau}
+{v}_{y_1}{v}_{y_3 x}-{v}_{y_1 x}{v}_{y_3}=0,
\\
{v}_{y_1 t_2} - {u}_{t_1} + {v}_{x y_1}{u}
- {v}_{y_1}{u}_{x}=0.
\eeaa

The considered example gives us some insight into 
the structure of general multidimensional quasiclassical
SDYM hierarchy for the Lie algebra of one-dimensional vector fields. We will now 
give a sketch of general picture, which will
be developed in more detail below.

For the case of $(M+2)$-dimensional hierarchy, $M\geqslant 3$,
corresponding to distribution of codimension $M$ (and,
respectively, equations of maximal dimensionality $M+2$)
we have $M-1$ infinite series of variables $t^k_n$,
$1\leqslant k \leqslant M-1$, $1\leqslant n <\infty$
and a variable $x$ (one-dimensional vector field variable). The hierarchy
is defined by linear equations
\bea
\p_{t^k_{n+1}}\Psi=(\lambda \p_{t^k_{n}}
-(\p_{t^k_{n}}v)\p_x)\Psi,
\quad 
1\leqslant k \leqslant M-1,
\quad 
1\leqslant n <\infty,
\label{NYM}
\eea 
corresponding to the basic vectors of the distribution
in recursive form. Equations (\ref{NYM}) have $M-1$
analytic (polynomial for finite number of times) solutions
\beaa 
\Psi^k=\sum_{n=0}^{\infty}t^k_{n+1}\lambda^n
\eeaa
and one solution of the form (\ref{PsiX}),
\beaa 
\Psi=x+\sum_{n=1}^\infty \Psi_n(\mathbf{t},x)\lambda^{-n}.
\eeaa
To obtain equations of maximal dimensionality $M+2$ in the 
framework of this hierarchy, it is possible to apply the scheme
of restriction to the submanifold used to derive system (\ref{5pair36}), which is easily generalized to the multidimensional case. The structure of the Lax pair
in this case is
\bea 
&&
\p_{t_1}\Psi=(\lambda \p_{y_1} - (\p_{y_1}v)\p_x)\Psi,
\nn
\\
&&
\p_{t_2}\Psi=\left(\sum_{n=1}^{M-2} \lambda^n\p_{\tau_n}
 + P_{M-3}\p_x\right)\Psi,
\label{multiLax}
\eea  
where we have $M+2$ variables $t_1$, $y_1$, $t_2$, $\tau_k$,
$1\leqslant k\leqslant M-2$, $x$, and $P_{M-3}$ is a polynomial
in $\lambda$ of the order $M-3$. Compatibility of this equations 
gives $(M+2)$-dimensional closed system for $v$ and coefficients of the polynomial $P_{M-3}$.
\section{Quasiclassical SDYM type hierarchies as 
a reduction of general
multidimensional case}
In this section we will demonstrate that 
quasiclassical SDYM type hierarchies
represent a simple and natural reduction of general
linearly degenerate 
multidimensional dispersionless hierarchy. We define this reduction
in terms of wave functions of polynomial distribution describing the hierarchy. We will introduce generating relations and the dressing scheme for the reduced hierarchy. The functional freedom in the
reduced dressing scheme (number of variables of the functions defining the dressing data) doesn't change in the process of reduction,
indicating that multidimensional  universal hierarchy is of the same
dimensionality as initial general case.
\subsection{Polynomial linearly degenerate
multidimensional dispersionless
hierarchy}
\label{SectionHierarchy}
Equations of the class we consider here are usually represented as commutation relations
for a pair of vector fields (Lax pair) with coefficients holomorphic
with respect to the spectral variable, they are usually
called dispersionless (or quasilinear) integrable equations.
In general, vector fields contain a derivative over spectral
variable (e.g., dispersionless KP equation).
A picture of general multidimensional dispersionless integral hierarchy,
starting from the properties of the wave functions,
generating relation in terms of differential $N$-form and Lax-Sato equations,
was developed in \cite{BDM}, \cite{LVB09}

In this work we consider Lax pairs 
with a spectral variable
entering vector fields only
as a parameter (there is no derivative over spectral variable in vector
fields). We will use the term linearly degenerate for this
case. Sometimes it is called weakly nonlinear, in twistor
theory the term hyper-CR case is also used. Typical representatives of this class are Pleban\'ski heavenly
equations and equation (\ref{Pavlov0}).

First, we will outline the picture of linearly degenerate
dispersionless hierarchy, using
the description given in \cite{LVB09} for the general dispersionless hierarchy,
starting from wave functions and generating differential $M$-form.

The involutive distribution of vector 
fields of codimension $M$ 
can be defined through the differential form 
\cite{BKgrass}, \cite{BK2014}
\beaa
\Omega _{M}=d\Psi ^{1}\wedge d\Psi ^{2}\wedge \dots d\Psi^{M},
\eeaa
where the functions $\Psi^{i}$ are level functions for integral surfaces of the distribution, i.e., integral surfaces are given by
intersection of level surfaces for these functions. These functions
are annulated by vector fields considered as differential operators,
and thus they are solutions of linear equations of
the hierarchy (`wave functions').
For the general hierarchy \cite{LVB09} 
the functions $\Psi^{i}$  depend on the
set of times $\mathbf{t}$ and a spectral parameter $\lambda$, differentials do not contain $d\lambda$,
and level functions (wave functions) are formal series in $\lambda$ 
\bea
&&
\Psi^k=\Psi^k_0+\wt\Psi^k
\nn 
\\&&
\Psi^k_0=\sum_{n=0}^\infty t^k_n \lambda^{n},\quad
\wt\Psi^k=\sum_{n=1}^\infty \Psi^k_n(\mathbf{t}^1,\dots,\mathbf{t}^{M})\lambda^{-n},
\label{levelform}
\eea
where $1\leqslant k\leqslant M$, depending on 
$M$ infinite sequences of independent variables
$\mathbf{t}^k=(t^k_0,\dots,t^k_n,\dots)$, $t^k_0=x_k$.

The hierarchy is generated by the relation
\bea
\left( J^{-1}d\Psi ^{1}\wedge d\Psi ^{2}\wedge \dots d\Psi
^{M}\right) _{-}=0
\label{Gen}
\eea
where $\left( {\cdots}\right) _{-}$ denotes 
the projection to the part of 
$\left({\cdots}\right)$ 
with negative powers in $\lambda$ 
(respectively  $\left( {\cdots}\right) _{+}$ projects
to nonnegative powers)
and
$
J=\det
(\p_{j}\Psi^{i})_{i,j=1,\dots,M}.
$
Reduction $J=1$ corresponds to volume-preserving (divergence-free
vector fields) case. Relation (\ref{Gen}) implies
Lax-Sato equations defining the dynamics of wave functions
over higher times, moreover, it is equivalent to the set of
Lax-Sato equations \cite{LVB09}.

Generating relation represents a polynomiality condition for the coefficients of the form 
\beaa   
\wt\Omega_M = J^{-1}d\Psi ^{1}\wedge d\Psi ^{2}\wedge \dots d\Psi^{M},
\eeaa   
leading to the polynomiality of integrable distribution corresponding
to this form.
This condition can be provided using Riemann-Hilbert  problem
(see below).

To construct polynomial vector fields belonging to the integrable
distribution corresponding to relation (\ref{Gen}), it is convenient
to use the following observation \cite{BDM}:
\begin{prop}
\label{prop}
For arbitrary vector field $\hat V$ in the space 
of variables
$\mathbf{t}$ with coefficients polynomial in $\lambda$
and $\mathbf{\Psi}=(\Psi^1,\dots,\Psi^{M})$, where functions $\Psi ^{i}$ of the form (\ref{levelform})
satisfy
generating relation (\ref{Gen}),
\beaa 
(\hat V \mathbf{\Psi})_+=\mathbf{0}
\implies
\hat V \mathbf{\Psi} =\mathbf{0}.
\eeaa  
\end{prop}
This
observation  gives an opportunity to construct polynomial
vector fields belonging to the integrable distribution
(linear operators of the hierarchy) in the spirit of
Manakov ring technique \cite{ZM85}, controlling only `singular' part 
of the result of action of the operator on wave function (here
it is a multicomponent function). First, we can introduce a basic set
of vector fields 
$\hat V^i_n=\lambda^n\p_i$, 
where $\p_i=\frac{\p ~}{\p x_i}$,
Then for
arbitrary derivative over higher time $\p^k_n=\frac{\p ~}{\p t^k_n}$,
$n\geqslant 1$, there exists a unique linear combination of basic
vector fields such that
\beaa 
(\p^k_n\Psi)_+=(\sum_{i,m} u^i_m(\mathbf{t})\lambda^m\p_i \Psi)_+,
\eeaa 
and, due to Prop.\ref{prop}, we get linear equations
of the hierarchy in the form 
\beaa 
\p^k_n\Psi=\sum_{i,m} u^i_m(\mathbf{t})\lambda^m\p_i \Psi.
\eeaa 

Introducing the Jacobian matrix
$$
(\text{Jac}_0)=\left(\frac{D(\Psi^1,\dots,\Psi^{M})}
{D{({x_1,\dots,x_{M}})}}\right),\quad \det(\text{Jac}_0)=J,
$$
it is possible to write the hierarchy in
the Lax-Sato form
\bea
&&
\partial^k_n\mathbf{\Psi}=\sum_{i=1}^{M}
\left((\text{Jac}_0)^{-1})_{ik} \lambda^n)\right)_+
{\partial_i}\mathbf{\Psi},\quad
1\leqslant k \leqslant M,
\label{genSato}
\eea
where $1\leqslant n < \infty$,
$\mathbf{\Psi}=(\Psi^1,\dots,\Psi^{M})$. 


First flows of the hierarchy 
(lowest level integrable distribution)
read
\bea
\partial^k_1\mathbf{\Psi}=(\lambda \partial_k-\sum_{p=1}^{M} 
(\partial_k u_p)\partial_p)\mathbf{\Psi},\quad 1\leqslant k\leqslant M,
\label{genlinear}
\eea
where 
$u_k=\Psi^k_1$, $1\leqslant k\leqslant M$.
A commutativity for any pair of vector fields 
(\ref{genlinear}) 
(e.g., with $\partial^k_1$ and $\partial^q_1$, $k\neq q$)
implies closed nonlinear 
(M+2)-dimensional  system of PDEs for the set of functions $u_k$, 
which can be written in the form
\beaa
&&
\partial^k_1\p_q\hat U-\partial^q_1\p_k\hat U+[\p_k \hat U,\p_q \hat U]=0
\eeaa
where $\hat U$ is a vector field, $\hat U=\sum_{p=1}^{M} u_p \p_p$. 

Higher flows of the hierarchy can be written also in remarkably simple recursive form
\bea
\partial^k_n\mathbf{\Psi}=(\lambda \partial^k_{n-1}-\sum_{p=1}^{M} 
(\partial^k_{n-1} u_p)\partial_p)\mathbf{\Psi},\quad 1\leqslant k\leqslant M,
\label{recursive}
\eea
which can be easily checked using Prop.\ref{prop}.
Formally commuting a pair of vector fields (\ref{recursive})
with distinct higher times,
we get (M+4)-dimensional quasiclassical
SDYM equation
\beaa
&&
\partial^k_n \p^q_{l-1}\hat U-\partial^q_l\p^k_{n-1}\hat U+
[\p^k_{n-1} \hat U,\p^q_{l-1} \hat U]=0
\eeaa
(compare to (\ref{quasiYM})),
reflecting the fact that our generic (M+2)-dimensional hierarchy
can be naturally immersed into a special (M+4)-dimensional hierarchy
(see below).

\subsubsection*{The dressing scheme}
It is 
easy to see that $\tilde\Omega_M$ is invariant under diffeomorphism
\beaa
(\Psi ^{1},\dots, \Psi^{M})\rightarrow 
\mathbf{F}(\lambda,\Psi ^{1},\dots, \Psi^{M})
\eeaa
Let us consider functions $\Psi ^{k}$ holomorphic inside and outside the unit circle (we denote components $\Psi ^{k}_\text{in}$ and $\Psi ^{k}_\text{out}$,  $\Psi ^{k}_\text{out}$ correspond to
the series introduced above),
having a discontinuity on it. If they satisfy a nonlinear vector
Riemann-Hilbert problem (nvRHp)
\bea
(\Psi ^{1},\dots, \Psi^{M})_\text{in} =
\mathbf{F}(\lambda,\Psi ^{1},\dots, \Psi^{M})_\text{out},
\label{dressing}
\eea
then the form $\tilde\Omega_M$ is holomorhic 
in all the complex plane. 

Thus nvRHp gives a tool to construct  forms
$\tilde\Omega_M$, generating
integrable distributions with holomorphic (meromorhic) coefficients and solutions to corresponding involution
equations. See \cite{BK}, \cite{BDM}, \cite{LVB09}
for more detail, nonlinear vector $\dbar$ problem can be 
also used in the dressing scheme.
\subsection{Description of the class of reductions}
In this work we consider a class of reductions of
general hierarchy (\ref{Gen}), (\ref{genSato})
characterised by the condition that $M-P$,
$1\leqslant P < M$, of the series $\Psi^k$
are equal to `vacuum' functions $\Psi^k_0=\sum_{n=0}^\infty t^k_n\lambda^n$ (for finite subsets of times they are
polynomial)
\be 
(\Psi^k)_-=0,\; \Psi^k=\Psi^k_0=\sum_{n=0}^\infty t^k_n\lambda^n,
\quad P< k\leqslant
M.
\label{reduction}
\ee 
Generating relation (\ref{Gen}) for the reduced hierarchy
takes the form
\beaa
\left( J^{-1}d\Psi ^{1}\wedge \dots d\Psi ^{P}\wedge 
\Psi^{P+1}_0
\wedge \dots d\Psi
^{M}_0 \right) _{-}=0,
\eeaa
where the Jacobian is effectively taken for the set of functions
$\Psi ^{1},\dots,\Psi ^{P}$.

Lax-Sato equations (\ref{genSato}) have rather special structure
in this case, taking into account that the only nonzero entries of last $M-P$ lines of the Jacobian
form the unity matrix.

Reduction (\ref{reduction}) is characterized rather simply
in terms of the dressing scheme (\ref{dressing}).
The dressing data for this scheme are represented by the
diffeomorphism $\mathbf{F}$. Let us take a diffeomorphism
that gives an identical transformation for last
$M-P$ components,
\beaa 
{F}^k(\lambda,\Psi ^{1},
\dots, \Psi^{M})=\Psi^k,
\quad P< k\leqslant M.
\eeaa 
Then the functions $\Psi^k$, $P<k\leqslant M$,
don't have a discontinuity
on the unit circle and are analytic in the complex plane,
thus leading us to reduction (\ref{reduction}).
The Riemann-Hilbert problem is then written for the first
$P$ functions $\Psi^k$, 
\beaa
\Psi ^{q}_\text{in} =
F^q(\lambda,\Psi ^{1},\dots,\Psi^P,
\Psi_0^{P+1},\dots, \Psi_0^{M})_\text{out},
\quad 1\leqslant q \leqslant P
\eeaa
The functional freedom of
the dressing data is $p$ functions of $M+1$ variables,
that indicates that reduced equations are generically
$M+2$-dimensional, like equations of unreduced hierarchy.

However, if we consider the hierarchy with complete initially
introduced set of times $\mathbf{t^1},\dots, \mathbf{t^M}$, 
we discover that the hierarchy contains subhierarchies with
different dimensionalities (different number of independent variables
in corresponding equations). Geometrically this is due to the
fact that some of the level functions $\Psi^i$ defining integral
surfaces for vector fields are rather special and depend 
only on some subset of the variables of the hierarchy (level
functions $\Psi^i=\Psi^i_0$). 
Thus, we can choose some subsets of
the variables of the hierarchy (finite or infinite), for
which the effective number of level functions is less than $M$,
then the codimension of corresponding distribution is
less than $M$, and dimensionality of corresponding equations of the hierarchy
(involutivity equations) is less than $M+2$.
It would be interesting to study geometry of distributions for
this hierarchy, which is probably described locally in terms of some special submanifolds of the Grassmannian
\cite{BKgrass}. However, in our examples we will try to choose
special subsets of times to get rid of this degeneracy and obtain $M+2$-dimensional equations.

The case $M=4$, $P=2$ after additional divergence-free reduction leads to the six-dimensional heavenly equation
\cite{PlebPrzan96}. For $P>2$, $M=P+2$ we get equation considered in \cite{BK}, \cite{ManSan2006}. We are planning to discuss these cases in more detail later.

The multidimensional reduced hierarchies considered in this work correspond to the choice $P=1$, i.e., we have
one nontrivial function $\Psi$ and several vacuum 
functions $\Psi^i_0$. Vector fields in this case are effectively one-dimensional, and corresponding hierarchies
can be considered as multidimensional extensions of
Shabat-Mart\'inez Alonso universal hierarchy \cite{UniSA}.
\subsection{Universal hierarchy}
Universal hierarchy of Shabat and Mart\'inez Alonso \cite{UniSA}
for `positive' times is a general hierarchy described in subsection  \ref{SectionHierarchy} 
for $M=1$. In this case we have one wave function $\Psi$,
\beaa 
\Psi=\Psi_0+\wt\Psi,
\qquad 
\Psi_0=\sum_{n=0}^\infty t_n \lambda^{n},\quad
\wt\Psi=\sum_{n=1}^\infty \Psi_n(\mathbf{t})\lambda^{-n},
\eeaa
generating equation (\ref{Gen}) acquires the form
\bea 
\left( \Psi_x^{-1}d\Psi 
\right) _{-}=0,
\label{GenUni}
\eea 
and Lax-Sato equations are
\bea
&&
\partial_n {\Psi}=
\left((\Psi_x)^{-1} \lambda^n\right)_+
{\partial_x}{\Psi},
\label{genSatoSA}
\eea
where $x=t_0$, $\partial_n=\frac{\p~}{\p t_n}$.
The first two flows 
\beaa
\p_y\Psi&=&(\lambda-v_x)\p_x,
\\
\p_t\Psi&=&(\lambda^2-v_x\lambda - v_y)\p_x,
\eeaa
where $y=t_1$, $t=t_2$, $v=\Psi_1(\mathbf{t})$,
give a Lax pair for (2+1)-dimensional equation
(\ref{Pavlov0}),
\beaa 
v_{xt}=v_{yy}+v_x v_{xy}- v_{xx}v_y,
\label{Pavlov}
\eeaa 
discussed in the Introduction.

The Riemann-Hilbert problem (\ref{dressing}) for the dressing scheme
associated with universal hierarchy looks like
\beaa 
\Psi_\text{in} =
{F}(\lambda,\Psi_\text{out}).
\eeaa
The properties of solutions and inverse problem for equation
(\ref{Pavlov0}) were recently studied in  \cite{GrinSan}.

Lax-Sato equations (\ref{genSatoSA}) can be represented
in recursive form (\ref{3pairRec}),
\bea 
(\partial_{n}-\lambda\p_{n-1}) {\Psi}
=-(\p_{n-1} v)\p_x \Psi,\quad n\geqslant 1.
\label{unirecursion}
\eea 
\subsection{Four-dimensional universal hierarchy}
Let us consider a reduction of the four-dimensional hierarchy
(Section \ref{SectionHierarchy}, $M=2$) characterised by the condition
$$
\Psi^2=\Psi^2_0,
$$
We drop higher times $t^1_n$, $n>1$  to get
rid of (2+1)-dimensional part of the hierarchy corresponding
to universal hierarchy for these times and
consider two wave functions
\beaa
\Psi=x+\sum_{n=1}^\infty \Psi_n(\mathbf{t})\lambda^{-n},
\quad 
\Psi^1=\sum_{n=0}^\infty t_{n+1}\lambda^n,
\eeaa
generating relation (\ref{Gen}) for this case reads
\bea 
(\Psi_x^{-1} d\Psi\wedge \sum_{n=0}^\infty\lambda^n dt_{n+1})_-=0.
\label{Gen4}
\eea 
Lax-Sato equations of reduced hierarchy can be extracted from
general Lax-Sato equations (\ref{genSato}) and have a rather special structure
\bea
&&
\partial_{n+1}{\Psi}=(\lambda^n\p_y-
\left((\Psi_x)^{-1}\Psi_y \lambda^n\right)_+
{\partial_x}){\Psi},
\label{genSato4}
\eea
where $y=t_1$. They can also be represented in a very simple
recursive form, implied by Prop. \ref{prop},
\bea 
(\partial_{n+1}-\lambda\p_n) {\Psi}
=-(\p_n v)\p_x \Psi, \quad n\geqslant 1.
\label{recur4}
\eea 
First two flows in recursive form
\beaa 
&&
(\partial_{t}-\lambda\p_y) {\Psi}
=-(v_y)\p_x \Psi
\\
&&
(\partial_{w}-\lambda\p_t) {\Psi}
=-(v_t)\p_x \Psi,
\eeaa 
where $t=t_2$, $w=t_3$,
give a Lax pair for a four-dimensional equation
\bea 
v_{wy}=v_{tt}+v_y v_{xt}- v_{yx}v_t.
\label{Pavlov4}
\eea
Riemann-Hilbert problem for the dressing scheme associated with this case 
is
\bea
\Psi_\text{in} =
{F}(\lambda,\Psi_\text{out},\sum_{n=0}^\infty\lambda^n t_{n+1}),
\label{dressingUni4}
\eea
and the dressing data is represented by the function 
of three variables $F(\lambda,X,Y)$, that corresponds to the
functional freedom for Cauchy problem for equation 
(\ref{Pavlov4}).
\subsection*{Immersion of universal hierarchy into four-dimensional
universal hierarchy}
There exists a natural immersion of universal hierarchy into four-dimensional
universal hierarchy. 
On the level of distributions it manifests itself in the fact that the basis
(\ref{unirecursion}) can be represented as the basis (\ref{recur4}) plus
one extra vector (for $n=0$). Codimension of distribution (\ref{unirecursion})
is 1, and codimension of distribution (\ref{recur4})  equals 2, respective
dimensionality of involutivity equations (equations of the hierarchy)  is 3 and 4.
Having solutions of equations of universal hierarchy depending on some
set of higher times, we obtain solutions of four-dimensional universal hierarchy.

This immersion has also a simple interpretation in terms of generating
equation and dressing scheme.
Let us have a solution of generating equation for universal hierarchy
(\ref{GenUni}),
\beaa
&&
\left( \Psi_x^{-1}d\Psi 
\right) _{-}=0,
\\
&&
\Psi=\sum_{n=0}^\infty \lambda^n t_n + \wt\Psi,
\quad \wt\Psi=\sum_{n=1}^\infty \Psi_n(\mathbf{t})\lambda^{-n},
\eeaa 
then the function $\Psi'=x+ \wt\Psi$ satisfies generating
equation (\ref{Gen4}) for four-dimensional universal hierarchy
\beaa  
({\Psi'}_x^{-1} d\Psi'\wedge \sum_{n=0}^\infty\lambda^n dt_{n+1})_-=0,
\eeaa
because, evidently,
\beaa 
({\Psi'}_x^{-1} d\Psi'\wedge \sum_{n=0}^\infty\lambda^n dt_{n+1})_-=
(\Psi_x^{-1} d\Psi\wedge \sum_{n=0}^\infty\lambda^n dt_{n+1})_-=0
\eeaa
In terms of the dressing scheme the immersion corresponds
to some special choice of the dressing data for the problem
(\ref{dressingUni4}), namely
\beaa 
F(\lambda,X,Y)=G(\lambda,X+\lambda Y),
\eeaa 
which allows to get (special) solutions of four-dimensional
universal hierarchy from solutions of universal hierarchy.
\subsection{Five-dimensional universal hierarchy}
Let us take $P=1$, $M=3$ and consider the series
\beaa
\Psi=x+\sum_{n=1}^\infty \Psi_n(\mathbf{t})\lambda^{-n},\; 
\Psi^1=\sum_{n=0}^\infty t^1_{n+1}\lambda^n,\;
\Psi^2=\sum_{n=0}^\infty t^2_{n+1}\lambda^n,
\eeaa
generating relation (\ref{Gen}) for this case reads
\beaa  
(\Psi_x^{-1} d\Psi\wedge 
\sum_{n=0}^\infty\lambda^n dt^1_{n+1}
\wedge
\sum_{n=0}^\infty\lambda^n dt^2_{n+1}
)_-=0,
\eeaa  
and Lax-Sato equations (\ref{genSato}) split
into two copies of Lax-Sato equations for the
four-dimensional universal hierarchy (\ref{genSato4}),
\bea
&&
\partial^1_{n+1}{\Psi}=(\lambda^n\p_{y_1}-
\left((\Psi_x)^{-1}\Psi_{y_1} \lambda^n\right)_+
{\partial_x}){\Psi},
\nn\\&&
\partial^2_{n+1}{\Psi}=(\lambda^n\p_{y_2}-
\left((\Psi_x)^{-1}\Psi_{y_2} \lambda^n\right)_+
{\partial_x}){\Psi},
\label{genSato5}
\eea
where 
$y_1=t^1_1$, $y_2=t^2_1$, $\p^k_n=\frac{\p ~}{\p t^k_n}$.
Taking first two flows 
of this hierarchy, we get linear equations (\ref{5pair})
and five-dimensional quasiclassical SDYM equation (\ref{5eq}).
Thus, equations (\ref{genSato5})  represent Lax-Sato equation
of the hierarchy associated with five-dimensional quasiclassical SDYM equation.

Solutions to the hierarchy can be constructed using
nonlinear Riemann-Hilbert problem 
\beaa
\Psi_\text{in} =
{F}(\lambda,\Psi_\text{out},\sum_{n=0}^\infty t^1_{n+1}\lambda^n, 
\sum_{n=0}^\infty t^2_{n+1}\lambda^n)
\eeaa
with the dressing data represented by the function of 
four variables.

The hierarchy can be also represented in recursive form
\beaa 
\p^1_{n+1}\Psi=(\lambda \p^1_{n}
-(\p^1_{n}v)\p_x)\Psi,
\\
\p^2_{n+1}\Psi=(\lambda \p^2_{n}
-(\p^2_{n}v)\p_x)\Psi,
\eeaa 
where $1\leqslant n < \infty$.
\subsection{Multidimensional reduced hierarchy}
Let us consider multidimensional case
$P=1$, $M>3$, corresponding to $M$ wave functions
(with only $P=1$ nontrivial)
\beaa 
\Psi=x+\sum_{n=1}^\infty \Psi_n(\mathbf{t},x)\lambda^{-n}
\eeaa
and
\beaa 
\Psi^k=\sum_{n=0}^{\infty}t^k_{n+1}\lambda^n, \quad
1\leqslant k \leqslant M-1.
\eeaa
Generating relation (\ref{Gen}) for this case reads
\beaa  
(\Psi_x^{-1} d\Psi\wedge 
\sum_{n=0}^\infty\lambda^n dt^1_{n+1}
\wedge
\dots 
\wedge
\sum_{n=0}^\infty\lambda^n dt^{M-1}_{n+1}
)_-=0,
\eeaa 
and Lax-Sato equations are constructed of the
same building blocks as in the previous case,
\bea 
\partial^k_{n+1}{\Psi}=(\lambda^n\p_{y_k}-
\left((\Psi_x)^{-1}\Psi_{y_k} \lambda^n\right)_+
{\partial_x}){\Psi},
\quad 
1\leqslant k \leqslant M-1,\; n\geqslant 1.
\label{LSmulti}
\eea
where $\p^k_n=\frac{\p ~}{\p t^k_n}$,  $y_k=t^k_1$.
Recursive form of the hierarchy is
\beaa 
\p^k_{n+1}\Psi=(\lambda \p^k_{n}
-(\p^k_{n}v)\p_x)\Psi, \quad n\geqslant 1.
\eeaa 

Solutions to the hierarchy can be constructed using
nonlinear Riemann-Hilbert problem 
\beaa 
\Psi_\text{in} =
{F}(\lambda,\Psi_\text{out},\sum_{n=0}^\infty t^1_{n+1}\lambda^n,
\dots,
\sum_{n=0}^\infty t^{M-1}_{n+1}\lambda^n)
\eeaa
with the dressing data represented by the function of 
$(M+1)$ variables, that indicates that maximal dimensionality
of equations of the hierarchy is $(M+2)$.

Each of Lax-Sato equations (\ref{LSmulti}) is three-dimensional. To extract $(M+2)$-dimensional equation
in the framework of this hierarchy, it is necessary
to perform a restriction to some submanifold of variables
to get rid of degeneracy. The geometric origin of this
degeneracy  was explained in subsection\ref{Subexample},
where we have demonstrated how to perform a restriction
to submanifold and obtain 6-dimensional equation for
the case $M=4$ (\ref{5pair3}). The scheme of restriction 
can be easily generalized to multidimensional case.
Let us consider a system of lowest order
Lax-Sato equations (\ref{LSmulti})
\bea 
\p_{t_k}\Psi=(\lambda \p_{y_k} - (\p_{y_k}v)\p_x)\Psi,
\quad 
1\leqslant k \leqslant M-1,
\label{LSmulti0}
\eea
with the set of variables $t_k=t^k_2$, $y_k=t^k_1$, $x$.
Codimension of distribution (\ref{LSmulti0})
is equal to $M$, which corresponds to the codimension
of the hierarchy.
Let us restrict this set of variables to $(M+2)$-dimensional
submanifold
\beaa
&&
y_2=t_3=\tau_1,
\\&&
\dots
\\
&&
y_{k+1}=t_{k+2}=(-1)^{k+1}\tau_{k},
\\&&
\dots 
\\
&&
y_{M-2}=t_{M-1}=(-1)^{M-2}\tau_{M-3},
\\
&&
y_{M-1}=(-1)^{M-1}\tau_{M-2}
\eeaa 
with the variables $x$, $t_1$, $t_2$, $y_1$,
$\tau_1,\dots,\tau_{M-2}$.
Restriction of the distribution corresponding to equations
(\ref{LSmulti0}) to the tangent space for this submanifold
gives a 2-dimensional distribution (Lax pair)
\beaa 
&&
\p_{t_1}\Psi=(\lambda \p_{y_1} - (\p_{y_1}v)\p_x)\Psi,
\nn
\\
&&
\p_{t_2}\Psi=\left(\sum_{n=1}^{M-2}\lambda^n\p_{\tau_n}
 + P_{M-3}\p_x\right)\Psi,
\eeaa  
where $P_{M-3}$ is a polynomial
in $\lambda$ of the order $M-3$. Compatibility of these equations
(commutativity of vector fields)
gives $(M+2)$-dimensional closed system for $v$ and coefficients of the polynomial $P_{M-3}$.
\section*{Acknowledgements}
The authors are grateful to A.B. Shabat, V.E. Adler, 
B. Krulikov, A.V. Mikhailov and M. Dunajski
for useful discussions.
This work was partially supported by the RAS Presidium
program `Nonlinear dynamics in mathematical and physical sciences' and by the RFBR grant 14-01-00389.


\end{document}